\documentclass[11pt,letterpaper]{article}

\usepackage[utf8]{inputenc}
\usepackage[margin=1.5cm]{geometry}
\usepackage{algorithm}
\usepackage{algpseudocode}
\usepackage{titlesec}
\usepackage{tabu}
\usepackage{enumitem}
\usepackage{amssymb}
\usepackage[dvipsnames]{xcolor}
\usepackage{graphicx}
\usepackage{caption}
\usepackage{subcaption}

\newlist{selectlist}{itemize}{2}
\setlist[selectlist]{label=$\square$,leftmargin=*,noitemsep,topsep=0pt}

\usepackage{lmodern}

\usepackage{hyperref}
\hypersetup{
    colorlinks=true,
    linkcolor=blue,
    filecolor=magenta,      
    urlcolor=blue,
}

\newcommand{\ve}[1]{\ensuremath{\mathbf{#1}}} 
\newcommand{\RN}{\ensuremath{\mathbb{R}}} 

\newcommand{\code}[1]{\texttt{#1}}

\titleformat{\section}[block]{\hspace{1em}\bfseries}{\thesection.}{0.5em}{} 
\titleformat{\subsection}[block]{\hspace{1em}}{\thesubsection}{0.5em}{}

\begin{document}

{\Large \textbf{\textit{Noise fingerprints in quantum computers: Machine learning software tools}}}
\vskip0.5cm
\noindent
\textbf{
Stefano Martina$^{1,2,\star}$, Stefano Gherardini$^{4,2}$, Lorenzo Buffoni$^{3}$, Filippo Caruso$^{1,2}$}\\ \\
$^{1}$ Department of Physics and Astronomy, University of Florence, Via Sansone 1, Sesto Fiorentino, I-50019, Italy.\\
$^{2}$ European Laboratory for Non-Linear Spectroscopy (LENS), University of Florence, Via Nello Carrara 1, Sesto Fiorentino, I-50019, Italy.\\
$^{3}$ Physics of Information and Quantum Technologies Group, Instituto de Telecomunica\c c\~oes, University of Lisbon, Av. Rovisco Pais, Lisbon, P-1049-001, Portugal.\\
$^{4}$ CNR-INO, Area Science Park, Strada Statale 14, Basovizza (TS), I-34149 \& Scuola Internazionale Superiore di Studi Avanzati (SISSA), Via Bonomea, 265, Trieste, I-34136, Italy.\\
$^{\star}$ Corresponding author: stefano.martina@unifi.it \\

\noindent
\textbf{Abstract}\\
In this paper we present the high-level functionalities of a quantum-classical machine learning software, whose purpose is to learn the main features (the fingerprint) of quantum noise sources affecting a quantum device, as a quantum computer. Specifically, the software architecture is designed to classify successfully (more than $99\%$ of accuracy) the noise fingerprints in different quantum devices with similar technical specifications, or distinct time-dependences of a noise fingerprint in single quantum machines.
\vskip0.5cm

\vskip0.5cm

\noindent
\textbf{Code metadata}\\

\noindent
\begin{tabular}{|p{6.5cm}|p{9.5cm}|}
\hline
Permanent link to code/repository used for this code version & \url{https://github.com/trianam/learningQuantumNoiseFingerprint} \\
\hline
 Permanent link to Reproducible Capsule & \url{https://codeocean.com/capsule/fa6e1d85-c99f-4a38-9c16-ac204da85040/} \\
\hline
Legal Code License   & GNU General Public License v3.0 \\
\hline
Code versioning system used & GIT \\
\hline
Software code languages, tools, and services used & python, IBM quantum services \\
\hline
Compilation requirements, operating environments \& dependencies & numpy, qiskit, qiskit\_terra, scikit\_learn \\
\hline
Support email for questions & \href{mailto:stefano.martina@unifi.it}{\nolinkurl{stefano.martina@unifi.it}} \\
\hline
\end{tabular}\\
\vskip0.5cm

\section{Introduction}

Quantum technologies are a fast developing scientific and industrial field \cite{DowlingPTRSA2003}. They have been already implemented in several different platforms, as for instance photonic circuits \cite{OBrienNatPhot2009,WangNatPhot2020}, but also Rydberg atoms \cite{AdamsReview2020}, superconducting devices \cite{devoret2004superconducting} and others. Likely, the most promising quantum technology is represented by quantum computers, i.e., quantum devices for quantum computing, among which it is worth mentioning superconducting circuits \cite{clarke2008superconducting,WuPRL2021}, trapped-ions quantum computers \cite{wineland2003quantum,PogorelovPRXQuantum2021}, photonic chips \cite{spring2013boson,metcalf2014quantum} and topological qubits \cite{freedman2003topological}. Both academic laboratories and industrial companies are devoting lots of effort and funding to boost research and technological improvements, towards the so-called \textit{quantum supremacy} \cite{AruteNature2019}, i.e., a quantum advantage to solve (numerical) problems that no classical computer will never solve. The actual drawback of these devices is the absence of a standard hardware (and thus even software) architecture, on which research activities may be jointly coordinated. For each of these platforms, indeed, ad hoc solutions are proposed and then realized, and this makes such a technologies still very expensive and incompatible from a device to another. 

However, in quantum computing, the main issue to be still solved is the unavoidable presence of external noise sources that dramatically limit the accuracy of quantum computations, as well as the large-scale realization of quantum circuits and algorithms. The negative impact of noise on quantum computing is so noticeable that the acronym \emph{Noisy Intermediate-Scale Quantum} (NISQ) technology has been recently introduced \cite{preskill2018quantum}. Furthermore, commercial quantum devices as for example the quantum computers by the companies Q-IBM\textsuperscript{\textregistered} \cite{ibmq} and Rigetti\textsuperscript{\textregistered} \cite{rigetti}, albeit they have been made available by anyone who creates a free account on their database, are not physically accessible and several specifications on the chip's parameters are not made public. 

In the paper \cite{martinaArXiv2021Learning}, we have recently observed on some IBM quantum computers that main features of the noise sources affecting the devices are specific of each single computer and have a clear time-dependence. For such a purpose, a \emph{testbed quantum circuit} -- composed by a fixed number of qubits -- is designed, then made run for a sufficient number of times and finally locally measured in correspondence of each qubit. From the measurements of the qubits (the measurement observable was the Pauli matrix $\sigma_z$), a set of measurement outcomes is recorded, collected, and then used to train a \emph{machine learning} (ML) algorithm \cite{BishopPRML2006,HastieESL2009}. However, it is worth noting that in \cite{martinaArXiv2021Learning} the features of the quantum noise are not reconstructed but just classified from a quantum device to another. Specifically, the classification task was successfully carried out by means of a \emph{support vector machine} (SVM) \cite{HastieESL2009,BishopPRML2006}, with a classification accuracy equal or greater than $99\%$. Hence, thanks to our procedure, one just needs to collect an informative statistics of quantum measurement outcomes (that are \emph{quantum data}) from the testbed quantum circuit, and subsequently train ML (classical) algorithms. In fact, no quantum noise modelling is required nor, in principle, the testbed circuit has to be controlled by time-dependent pulses \cite{MuellerArXiv2020}. Also for these reasons, the use of a ML technique is the most natural choice to perform classification, since it naturally provides a black-box model with predictive outcomes. In this regard, we recall that in the current literature ML has been already adopted to distinguish open quantum dynamics \cite{YoussryArXiv,luchnikov2020machine,fanchini2020estimating} and to perform quantum sensing tasks \cite{niu2019learning,HarperNatPhys2020,MartinaArXiv2021,wise2021using}, as for example the learning and classification of non-Markovian noise \cite{niu2019learning,MartinaArXiv2021} or the detection of qubits correlations \cite{HarperNatPhys2020}.   

\begin{figure}[t!]
    \centering
    \includegraphics[width=0.9\textwidth]{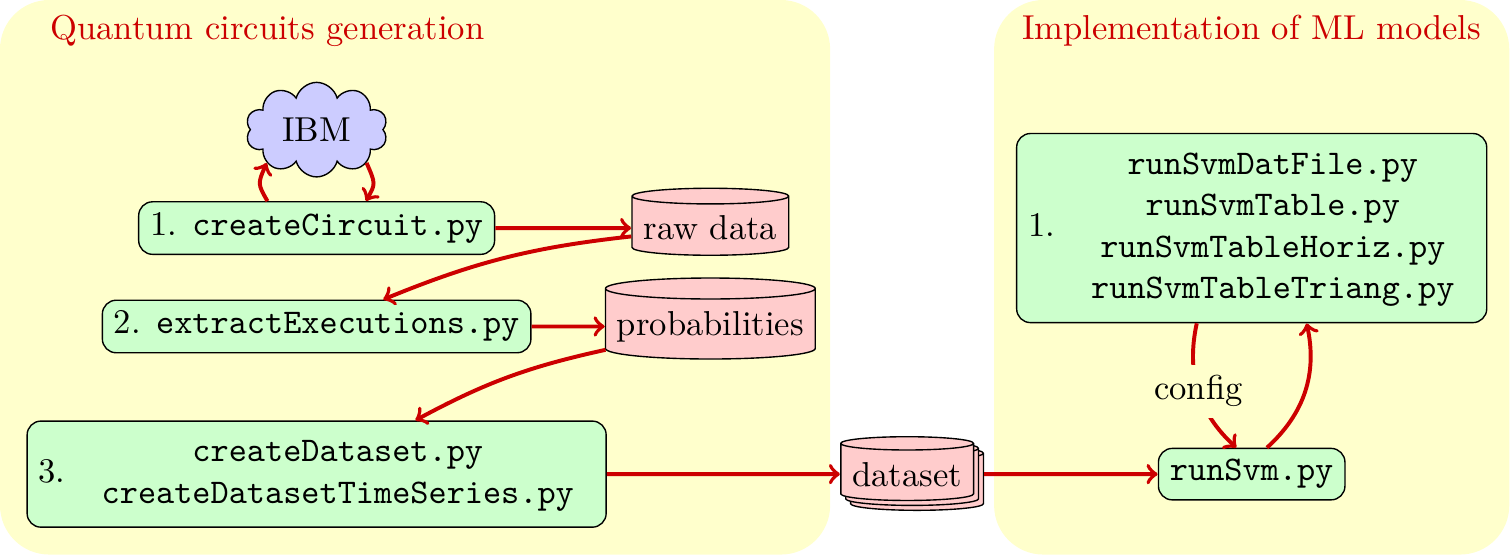}
    \caption{Pictorial figure showing the structure of the software architecture and how its different parts depend each other. On the left, one can observe the part of the software that is designed for the generation of the testbed quantum circuit. Specifically, \code{createCircuit.py} is used to launch the quantum circuit on the IBM cloud services with the aim to get the raw data from the measurement procedure in each execution of the circuit. The file \code{extractExecutions.py} computes the probabilities to get the measurement outcomes from the execution of the testbed quantum circuit. Either \code{createDataset.py} and \code{createDatasetTimeSeries.py} is used to pack in datasets the measurement outcomes probabilities. The former creates datasets with data collected on two or more machines, while the latter collects data taken on the same machine but at different times. On the right side of the figure, we represent the workflow for the training of ML methods. Specifically, the files \code{runSvmDatFile.py}, \code{runSvmTable.py}, \code{runSvmTableHoriz.py} and \code{runSvmTableTriang.py} are employed to generate the output data that we have shown in the tables and plots in \cite{martinaArXiv2021Learning}. All these scripts call the function \code{runSVM} in the file \code{runSvm.py} that contains the main code for the definition, training and evaluation of the SVM models. Note that the function is called with one of the configuration names (\code{config} in the picture) that is listed in the config file \code{configurations.py}. The configuration name denotes what is the generated dataset that is used for the training of the ML models.}
    \label{fig:functionalGraph}
\end{figure}
As depicted in Fig.\,\ref{fig:functionalGraph}, our software architecture adopted in \cite{martinaArXiv2021Learning} has two distinct parts: The one on the left of the figure generates the testbed quantum circuit (see Sec.\,\ref{sec:circuit_generation}), while the other is designed for the implementation of the ML models that classify quantum noise fingerprints (refer to Sec.\,\ref{sec:ML_models}).

\section{Testbed quantum circuit}\label{sec:circuit_generation}

For our experiments of quantum noise classification in \cite{martinaArXiv2021Learning}, we made use of the IBM Quantum cloud services to remotely run quantum circuits on different machines. In particular, to interact with the cloud services, one can use the Qiskit sdk \cite{qiskit} that is an open-source python package, useful both to \textit{simulate} quantum dynamics and to \textit{program} a given set of operations on a real quantum computer. Currently, one has at disposal up to $11$ superconducting quantum computers, ranging from a single qubit up to $15$ qubits, with different topology and calibration routines. For all the available devices and their specifications, we direct the reader to the IBM documentation \cite{ibmq}.


\begin{algorithm}[t!]
\caption{Generation of the testbed quantum circuit (baseline version)}\label{alg:cap}
\begin{algorithmic}
\Require IBM-Q backend (specific device to fingerprint)
\Ensure $|0\rangle_i$ $\forall i \in 0,...,3$
\For{number of repetitions}
\State $0 \gets H$ 
\Comment Hadamard gate on the $0^{th}$ qubit
\State $1 \gets H$
\State ${\rm CNOT}(0 \to 2)$ \Comment Controlled NOT gate on the $2^{nd}$ qubit conditioned on the qubit $0$
\State ${\rm CNOT}(1 \to 3)$
\State $0 \gets X$ 
\Comment X gate on the $0^{th}$ qubit
\State $1 \gets X$
\State ${\rm Toffoli}(0,1 \to 2)$
\Comment Toffoli gate on the $2^{nd}$ qubit conditioned on the qubits $0,1$
\EndFor
\State Measure(2)
\Comment Projective measurements of the $i^{th}$ qubit
\State Measure(3)
\State\Return 1000 shots from the measurements
\end{algorithmic}
\end{algorithm}
\begin{figure}[h!]
    \begin{subfigure}[b]{0.5\textwidth}
         \centering
         \includegraphics[height=4.25cm]{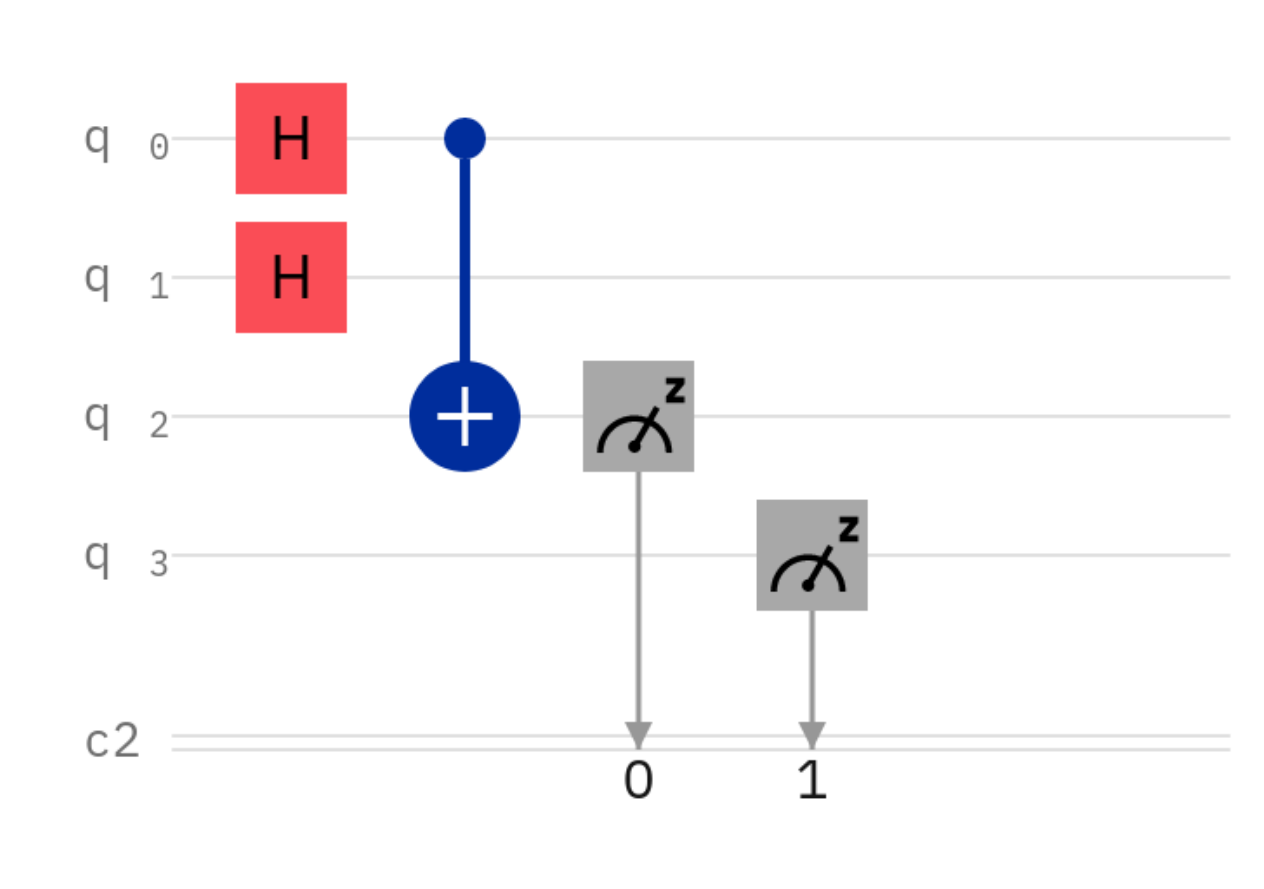}
         \caption{}
         \label{fig:circuit1}
    \end{subfigure}
    \begin{subfigure}[b]{0.5\textwidth}
         \includegraphics[height=4.25cm]{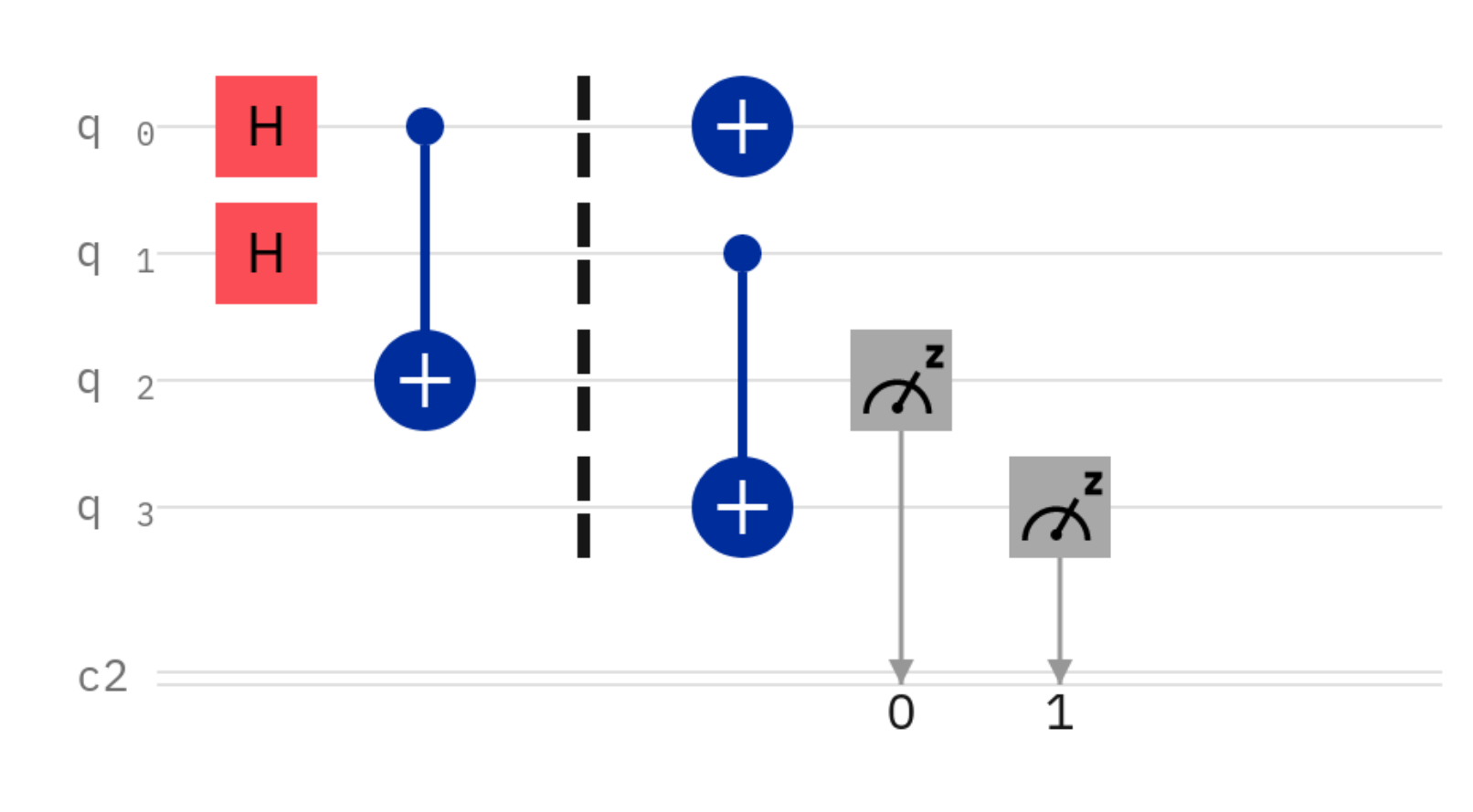}
         \caption{}
         \label{fig:circuit2}
    \end{subfigure}
    \begin{subfigure}[b]{\textwidth}
         \centering
         \includegraphics[height=4.25cm]{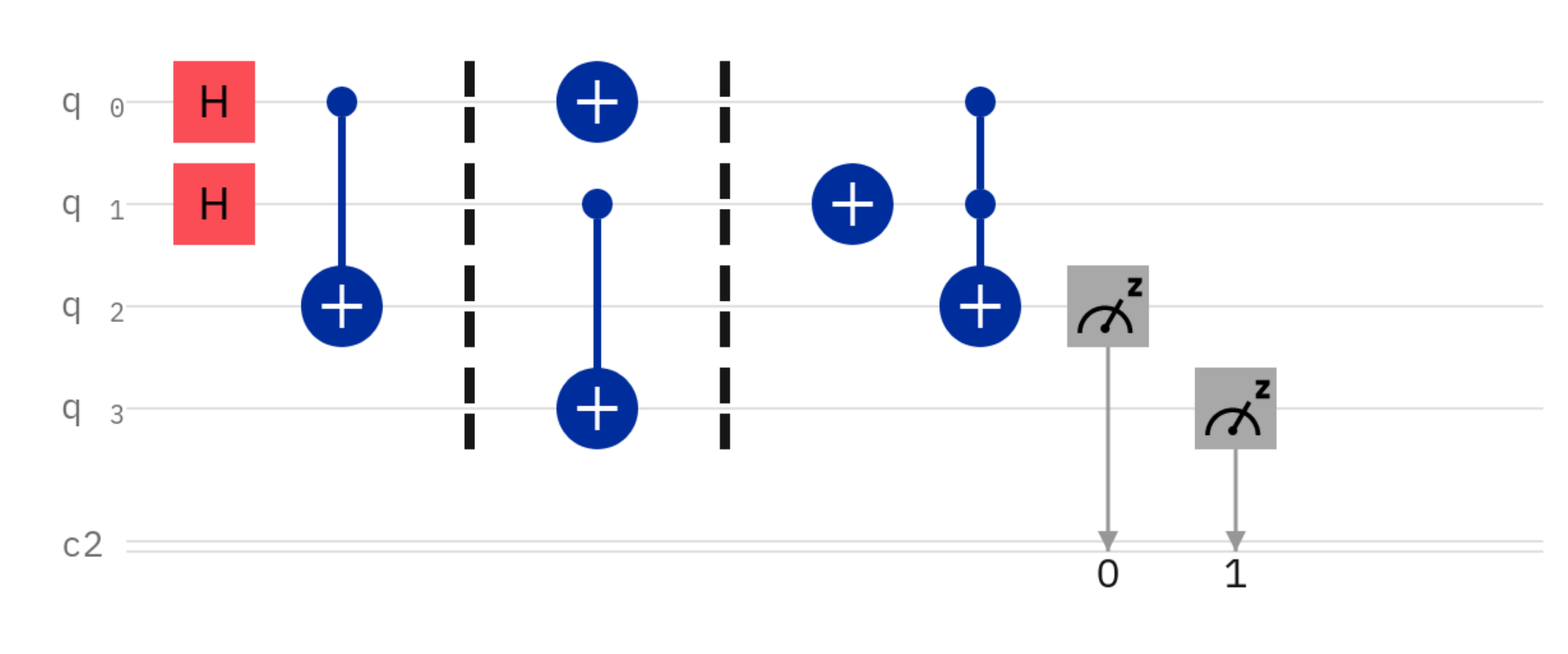}
         \caption{}
         \label{fig:circuit3}
    \end{subfigure}
    \begin{subfigure}[b]{\textwidth}
         \centering
         \includegraphics[height=4.25cm]{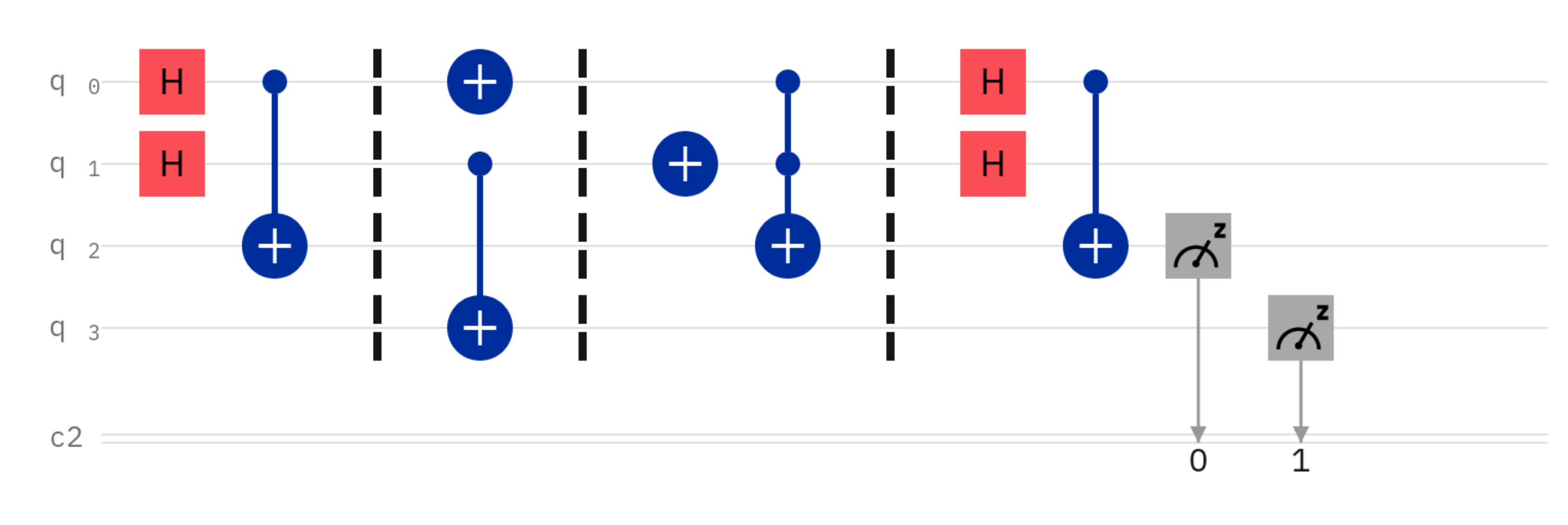}
         \caption{}
         \label{fig:circuit4}
    \end{subfigure}
    \caption{Pictorial representation of the first 4 measurement steps (the panels of the figure) applied to the testbed quantum circuit designed in Ref.\,\cite{martinaArXiv2021Learning}. The 4 panels have to be read from left to right, and from top to bottom. The baseline version of the testbed quantum circuit (provided by Algorithm 1) is the one depicted in panel (c), in correspondence of the third measurement step. The full set of measurement outcomes is obtained by repeating 3 times the baseline circuit and then performing a total of 9 measurement steps, each of them acting on the qubits 2 and 3. The outcomes collected in the 9 measurement steps come from executing incrementally the testbed quantum circuit in different runs, where the qubits 2 and 3 are measured only at the end of the implemented circuits.}
    \label{fig:circuit}
\end{figure}

In Algorithm 1 we provide the pseudo-code for the generation of the testbed quantum circuit in its baseline version (see also panel (c) in Fig.\,\ref{fig:circuit} for a pictorial representation of the circuit) realized in \cite{martinaArXiv2021Learning} to carry out the classification of noise fingerprints.  
For the quantum computation in the aforementioned circuit, we made use of standard gates whose mathematical definitions 
is given in terms of matrices that one can easily find in quantum computing textbooks \cite{nielsen2002quantum}. The baseline version of the testbed quantum circuit is repeated 3 times overall for a total of 9 measurements (also denoted as measurement steps) of both qubits 2 and 3. Indeed, operationally, the 9 measurement steps are not performed all in the same run (i.e., sequentially), but on consecutive runs by implementing incremental parts of the quantum circuit. To clarify this aspect as much as possible, in Fig.\,\ref{fig:circuit} we have represented pictorially the first 4 measurement steps of the testbed quantum circuit, whose baseline version (returned by Algorithm 1) is depicted in the panel (c) of the figure. Specifically, first we execute the part of the circuit that is obtained by cutting the testbed quantum circuit after the first measurement step, i.e., after the measurement of qubit 2 and 3 following the Hadamard gates on the $0^{th}$ and $1^{th}$ qubits and the CNOT gate from qubit $0$ to $2$). Then, the measurement outcomes are recorded. Subsequently, we execute part of the testbed quantum circuit until the second measurement step (measurements of qubits 2 and 3 included), thus by also taking into account the $X$-gate on the $0^{th}$ qubit and the CNOT gate from qubit $1$ to $3$, and again we record the measurement outcomes. The procedure is then repeated for all the 9 measurement steps. Before proceeding, it is worth stressing that, for each implemented quantum circuit, the measurements of the qubits 2 and 3 are performed only at the end of the circuits. 

Overall, in \cite{martinaArXiv2021Learning}, several experiments have been conducted on different IBM chips that have different physical specifications, as the architecture of the qubits or the quantum volume \cite{cross2019validating}. 
Some quantum machines, indeed, are inherently noisier than other, and even single qubits inside a machine can have a distinctive noise profile. All these peculiar differences in noise and topology contribute to the fingerprint that we aim to classify using our ML method. 

\subsection{Data acquisition}

The pipeline designed for the creation of the dataset, set as input of the ML models, is constituted of several scripts that can be customized according to the needs of the user. First, for each implemented quantum circuit, the script \code{createCircuit.py} adopts Qiskit to interact with the IBM quantum services for the measurement of a predefined number of circuit executions. Specifically, such a script is parameterized to launch the runs of the circuits on several quantum machines with a specific amount of parallel tasks. The runs are executed in two different modalities that generates the datasets that we called FAST and SLOW in \cite{martinaArXiv2021Learning}. In the former dataset, the aim was to collect as many runs as possible in the shortest time interval. For this purpose, the script launches 20 parallel processes, each of them adds to the IBM queue a predefined number of runs with 8000 execution-shots. After that, each batch of 8000 execution-shots is split into 8 batches of 1000 shots that are then employed to compute the outcomes' probabilities. 
Instead, for the dataset named SLOW, we collect a sequence of measurement outcomes that are uniformly distributed over time. To obtain such dataset, the script launches only one run at a time with 1000 execution-shots and waits two minutes from a run to another. 

The second script in the pipeline is \code{extractExecutions.py}, whose objective is to compute the probabilities of the measurement outcomes from the raw data returned by the calls to the IBM quantum services (this is ensured by the previous script). After that, either \code{createDataset.py} and \code{createDatasetTimeSeries.py} pack the probability distributions in datasets. The difference between such scripts is the following. The former builds binary or multiclass classification dataset using data from at least two quantum machines, while the latter builds classification datasets with data collected in a single machine and labelled by the time interval in which the testbed quantum circuit is executed.

In the github repository at \url{https://github.com/trianam/learningQuantumNoiseFingerprint} and on CodeOcean at \url{https://codeocean.com/capsule/fa6e1d85-c99f-4a38-9c16-ac204da85040/}, we release the source code of all the scripts and all the data obtained from the execution of \code{createCircuit.py} on each quantum machine we employed.

\section{Machine Learning models}\label{sec:ML_models}

\subsection{Support Vector Machine}

In \cite{martinaArXiv2021Learning} we have successfully classified the noise fingerprints on several IBM quantum computers by training Support Vector Machine models \cite{HastieESL2009,BishopPRML2006}. SVM is a machine learning technique that is usually used to solve binary classification tasks. Generally speaking, a SVM model is trained on a dataset composed of pairs $(\ve{x}_i,y_i)$, where $\ve{x}_i$ are points in a certain space $\RN^n$ of dimension $n$, and $y_i$ is equal to $1$ or $-1$ depending the corresponding point belongs to one or the other class. The Support Vector Machine is trained to find the hyperplane that divides the space representation of the two classes, by ensuring the maximum distance from the points. When the points of the two classes are not linearly separable, a common solution is to resort to the so-called ``kernel trick'', i.e., the points $\ve{x}_i$ are mapped to a larger dimension space until they become linearly separable. The most common kernels are polynomial functions with varying degree number and the so-called Radial Basis Functions (RBF) \cite{HastieESL2009, BishopPRML2006}. Finally, SVM models can also be extended to multiclass classification tasks using the strategies One-Versus-All (OVA), or One-Versus-One (OVO) \cite{HastieESL2009, BishopPRML2006}.

In our work, to implement and train the SVM models, we leveraged the \emph{scikit-learn} python library \cite{scikit-learn}.

\subsection{Data interpretation}

The code that implements and train the SVM is defined by the functions in the file \code{runSVM.py}. Specifically, the main function is called \code{runSVM}: it requires a configuration object that \emph{(i)} identifies what model has to be used, and \emph{(ii)} set optional arguments to tune the number of hyperparameters (\code{mask}) and to control if the method is verbose (\code{verbose}) and if the results have to be written in an output file (\code{writeToFile}). Practically, the function \code{runSVM} first calls \code{extractData} whose purpose is to load the dataset file, extract the data in the desired time steps and split them in \emph{training}, \emph{validation} and \emph{test} sets. After that, \code{runSVM} proceeds to train a set of possible SVM models on the training set, by then evaluating them on the validation set and computing on the test set the resulting accuracy of the model that performed better on the validation set. The possible models that can be employed are: \emph{(i)} Standard linear SVM (using two different libraries), \emph{(ii)} SVM with \emph{polynomial} kernel with degree from 2 to 4, and \emph{(iii)} SVM with RBF kernel.

The file \code{runSVM.py} is provided with a \code{main} method. Thus, it can be directly called as a script using the configuration name as argument. We have also designed some useful methods that call \code{runSVM}, build directly the latex table with the results and calculate the points for the figures shown in \cite{martinaArXiv2021Learning}.

\section{Impacts}

In this paper, we have explained in great detail the software architecture of the ML method, introduced in \cite{martinaArXiv2021Learning}, to carry out quantum noise classification. Such a tools are intended to be applied to quantum technologies, as e.g., quantum computers. 

The main impact of our software lies in its ability in classifying the fingerprint left by quantum noise sources on devices that have identical technical specifications and are thus expected to provide the same outcomes. Unfortunately, in quantum machines, the influence of the environment is so relevant that different noise fingerprints can be identified depending on the type of quantum computer (as previously explained, quantum computers can differ, e.g., on the number of qubits and/or the quantum volume), on the time period in which the single machine has worked, and on environmental changes mainly due to temperature fluctuations. However, thanks to our quantum-classical machine learning method, one can \emph{(i)} distinguish the noise fingerprints in different quantum devices; \emph{(ii)} classify the noise fingerprint on the same quantum devices but in different times; \emph{(iii)} learn if and how a given noise fingerprint changes over time.

In \cite{martinaArXiv2021Learning} our method is proved to be very \emph{accurate} (more than $99\%$ of effectiveness) in classifying a clear machine-related noise fingerprint in each of the analysed IBM quantum computers, and even \emph{robust} since any noise fingerprint is highly predictable over time in windows of consecutive runs. Also an evident time-dependence of the noise fingerprints has been classified, by observing changes over time after few hours from the first execution of the testbed quantum circuit.

Another important feature of our software architecture is that the ML models do not require a complete set of measurement outcomes as input data, but conversely the outcomes from a sequence of repeated measurements of a single observable. For an example, for the experiments in \cite{martinaArXiv2021Learning}, the chosen observable was the tensor product of $\sigma_z$ Pauli matrices locally applied on each qubit of the testbed circuit. Furthermore, the proposed method is able to distinguish and classify noise fingerprints, even without knowing the microscopic model that describes the (real or effective) interaction between the device and the external noise fields. This important aspect allows the user to employ our quantum-classical machine learning algorithm to classify the noise fingerprints of even \emph{inaccessible} quantum machines.

\subsection{Applications}

The experimental evidences in \cite{martinaArXiv2021Learning} lead us to conclude that different quantum devices exhibit distinctive, and thus distinguishable, noise fingerprints that one can classify and predict. Therefore, in principle, our method could be adopted \emph{to identify} from which specific quantum device certain data (a collection of measurement outcomes) are generated, just looking at the noise fingerprint of the device. Moreover, the proposed solution might be employed \emph{to certify} the time-scheduling in which a given quantum computation is executed. Both these applications are expected to play a key role for diagnostics purposes -- especially in all those contexts where quantum computations cannot be error-corrected \cite{deutsch2020harnessing} -- and to accomplish benchmarking and certification \cite{WrightNatureComm2019,eisert2020quantum} of quantum noise sources within a default error threshold.

\subsection{Outlook}

The proposed methodology, aimed to learn the noise fingerprint of quantum devices from time-ordered measurements of a testbed quantum circuit, may be in principle applied to any quantum devices, and thus not only to the IBM quantum computers as done in \cite{martinaArXiv2021Learning}. The possibility to predict on which device, and at which time, a given quantum operation (even time-varying) has been executed is expected to help the mitigation of quantum computational errors (e.g., by means of calibration routines), and to assist the application of ad-hoc error corrections.

Furthermore, instead of SVMs, one could employ deep learning techniques, as for example Recurrent Neural Networks (RNN) \cite{BishopPRML2006,GoodfellowDL2016,schmidhuber2015deep}, to make more efficient the classification of quantum noise fingerprints. In such a case, the software architecture should be modified a bit, but not in the part that concerns the generation of the quantum data placed in input to the ML model. What should be different, indeed, is the way the input data would be processed.

Finally, we are also confident that, thanks to specific modifications, it is possible to carry out even the reconstruction of some quantum noise features. However, for such a purpose, a minimal knowledge of the way noise sources affect the quantum device under investigation will be required.  

\section*{Acknowledgements}

We acknowledge the access to advanced services provided by the IBM Quantum Researchers Program.\\
This work was financially supported from Fondazione CR Firenze through the project QUANTUM-AI, from University of Florence through the project Q-CODYCES, and from the European Union’s Horizon 2020 research and innovation programme under FET-OPEN Grant Agreement No.\,828946 (PATHOS).

\bibliographystyle{unsrt}
\bibliography{quantum,ml}

\end{document}